\documentclass{hotnets16}
\pdfoutput=1

\usepackage{times}  
\usepackage{epsfig}
\usepackage[TABBOTCAP]{subfigure}
\usepackage{tabularx}
\usepackage{graphicx} 
\usepackage{color}
\usepackage{xspace}
\usepackage{thumbpdf}
\usepackage{listings}
\usepackage{verbatim}
\usepackage{hyperref}
\usepackage{booktabs}
\usepackage{colortbl}
\usepackage{multirow}
\usepackage{diagbox}
\usepackage[disable]{todonotes}

\hypersetup{pdfstartview=FitH,pdfpagelayout=SinglePage}

\newcommand{\xxxTitle}[1]{\mbox{C-Share}}
\newcommand{\xxx}[1]{\emph{\mbox{C-Share}}}
\newcommand{\ignore}[1]{}
\newcounter{commentNumberI}
\newcommand\Mark[1]{\textsuperscript#1}

\newcommand{\new}[1]{\textcolor{blue}{#1}}
\newcommand{\old}[1]{\textcolor{red}{#1}}

\newcommand{\SV}[1]{\addtocounter{commentNumberI}{1} \todo[inline,color=blue!20]{\textbf{(\arabic{commentNumberI}.)} SV: #1}}

\newcommand{\TBD}[1]{\addtocounter{commentNumberI}{1}
\todo[inline,color=green!20]{\textbf{(\arabic{commentNumberI}.)} YBI: #1}}

\begin{document}

\conferenceinfo{HotNets 2016} {}
\CopyrightYear{2016}
\crdata{X}
\date{}

%%%%%%%%%%%% THIS IS WHERE WE PUT IN THE TITLE AND AUTHORS %%%%%%%%%%%%
\sloppy

\title{C-Share: Optical Circuits Sharing for Software-Defined Data-Centers}

% \author{Anonymous}
%\ignore{
\numberofauthors{1}
\author{Yaniv Ben-Itzhak\Mark{1}, Cosmin 
Caba\Mark{2}, Liran Schour\Mark{1}, Shay Vargaftik\Mark{1}\Mark{3}\\
\hspace{-0.5cm} \affaddr{\Mark{1}IBM Research Lab, Haifa, Israel} \hspace{1cm}
\affaddr{\Mark{2}DTU Fotonik}
\hspace{1cm}
\affaddr{\Mark{3}Technion}\\ 
\hspace{-0.5cm} \email{\{yanivb, lirans\}@il.ibm.com} \quad
\email{cosm@fotonik.dtu.dk} \quad
% \email{lirans@il.ibm.com} \quad
\email{shayvar@tx.technion.ac.il}
}
%}
\maketitle

%\thispagestyle{empty}

%%%%%%%%%%%%%  ABSTRACT GOES HERE %%%%%%%%%%%%%%
\subsection*{Abstract}
Integrating optical circuit switches in data-centers is an on-going research challenge.
In recent years, state-of-the-art solutions introduce hybrid packet/circuit architectures for different optical circuit switch technologies, control techniques, and traffic rerouting methods.
These solutions are based on separated packet and circuit planes which do not have the ability to utilize an optical circuit with flows that do not arrive from or delivered to switches directly connected to the circuit's end-points. 
Moreover, current SDN-based elephant flow rerouting methods require a forwarding rule for each flow, which raise scalability issues. 
In this paper, we present \xxx{} -- a practical, scalable SDN-based circuit sharing solution for data center networks. \xxx{} inherently enable elephant flows to share optical circuits by exploiting a flat upper tier network topology. \xxx{} is based on a scalable and decoupled SDN-based elephant flow rerouting method comprised of elephant flow detection, tagging and identification, which is utilized by using a prevalent network sampling method (e.g., sFlow). \xxx{} requires only a single OpenFlow rule for each optical circuit, and therefore significantly reduces the required OpenFlow rule entry footprint and setup rule rate. It also mitigates the OpenFlow outbound latency for subsequent elephant flows.  We implement a proof-of-concept system for \xxx{} based on Mininet, and test the scalability of \xxx{} by using an event driven simulation. Our results show a consistent increase in the mice/elephant flow separation in the network which, in turn, improves both network throughput and flow completion time.

% use: http://www.acm.org/about/class/ccs98-html
%\category{C.2.0}{Computer Communication Networks}{General}[Data communications)]
%\category{C.2.3}{Computer Communication Networks}{Network Operations}
%A category including the fourth, optional field follows...
%\category{D.2.8}{Software Engineering}{Metrics}[complexity measures, performance measures]
% \category{C.2}{Network Architecture and Design}{Circuit-switching networks}
% \terms{}
% \keywords{Packet/Circuit Networks, Optical Switches}

% \begin{figure}
% \centering
% \epsfig{file=fly.eps}
% \caption{A sample black and white graphic (.eps format).}
% \end{figure}

\section{Introduction} \label{sec:introduction}

\begin{figure*}[t!]
	\centering{\includegraphics[width=0.9\linewidth]{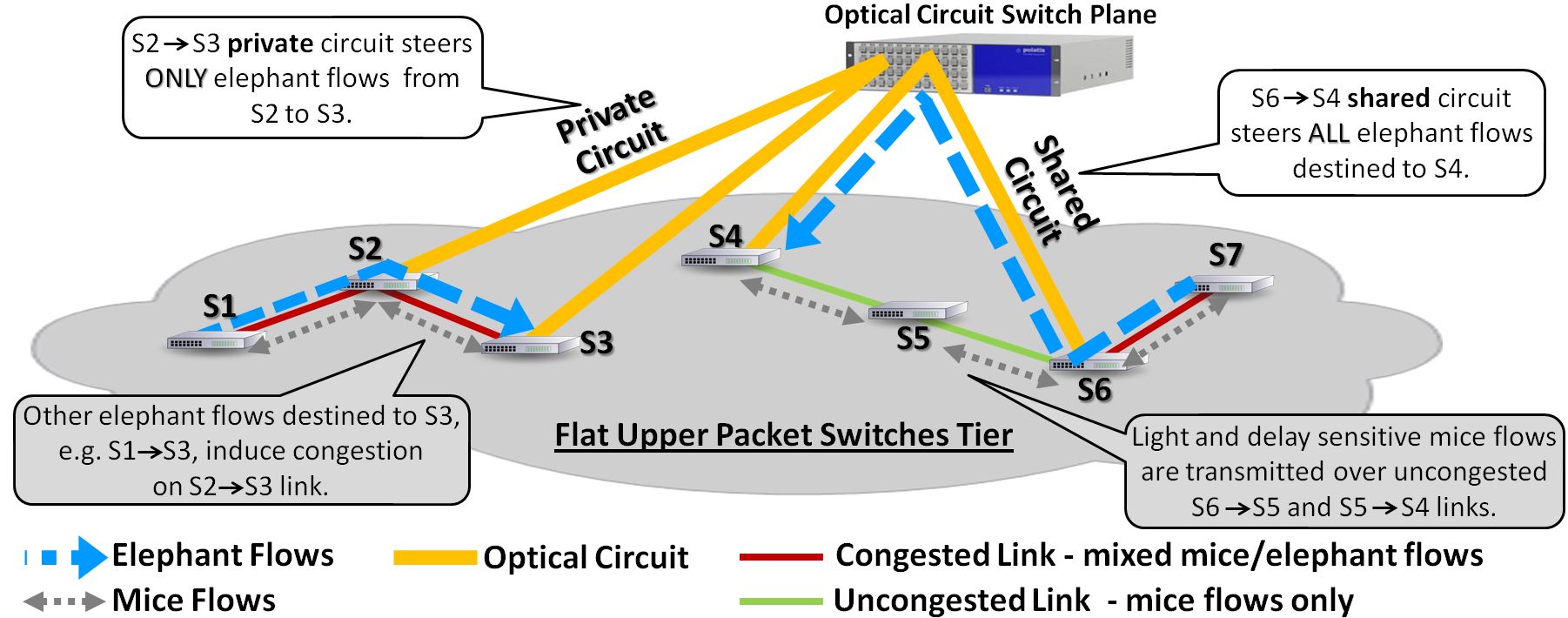}}
	\vspace{-0.3cm}
	\caption{The \xxx{} topology concept. \textit{Private} optical circuit results in inefficient mice/elephant flow separation over $S2$$\rightarrow$$S3$  link. On the other hand, a \textit{shared} optical circuit reduces the load over $S6$$\rightarrow$$S5$ and $S5$$\rightarrow$$S4$ links by better mice/elephant flow separation.}
    \vspace{-0.4cm}
	\label{fig:topology}
%    \vspace{-0.5cm}
\end{figure*}

In recent years, optical circuit switching has emerged as a promising solution for scaling data center networks. Current optical-circuit-switch/electrical-packet-switch (referred to as OCS/EPS) solutions, e.g. \cite{calient,farrington2011helios,REACToR,wang2011cThrough}, are based on separated OCS and EPS planes, employing the OCS for high-bandwidth, slowly varying, and long-lived flows (\textit{elephant flows}), and the EPS for fast varying and short-lived flows (\textit{mice flows}). Accordingly, each solution presents a method for detecting and rerouting elephant flows.

% drawbacks of current hybrid OCS/EPS? just the ones which relevant for CiShare ... look at related work
% The cons of current approaches: coupled, scalability (footprint), slow response for elephant flows - outbound latency, not SDN-based networks, sharing - later? (feature)

% scalability + coupling disadvantages
In the following we explain the lack of mice/elephant flow separation and scalability issues in current solutions. \quad\,\,
% optical circuit sharing ...
First, OCS can create low-latency high-bandwidth circuits\footnote{ In this paper we use \textit{circuit} and \textit{optical circuit} interchangeably.} using a relatively slow reconfigurable cross-board. 
OCS reconfiguration penalty, which is the time to establish a circuit, is tens of $\mu s$ for 2D MEMS wavelength selective switches, e.g., \cite{REACToR,mordia}\ignore{,legtchenko2016xfabric}, and tens of $ms$ for 3D MEMS optical circuit switches, e.g., \cite{calient,polatis,OSA,farrington2011helios,wang2011cThrough}.
Despite this penalty, previous solutions utilize a given optical circuit by transmitting only elephant flows that arrive from and delivered to switches directly connected to the optical circuit's end-points -- referred to as a \textit{private} circuit. 
Therefore, other elephant flows that are not assigned to an optical circuit are transmitted through the EPS plane. 
These elephant flows are usually high persistent TCP flows, which tend to fill the network buffers end-to-end. In turn, both elephant and mice flows that share these buffers are introduced with a non-trivial queuing delay. Therefore, delay sensitive mice flows and especially \textit{coflows}\footnote{ Collection of flows with a shared completion time that depends on completion time of the last-flow.}
~\cite{chowdhury2012coflow,chowdhury2014efficient,qiu2015minimizing,zhao2015rapier}
\TBD{add refs...}, are adversely affected.
% Therefore, further separation of mice and elephant flows is essential.

Second, state-of-the-art-solutions, e.g. \cite{calient,farrington2011helios}, introduce a coupled architecture in which both the detection and rerouting of elephant flows are employed over the switches directly connected to the OCS plane.
In particular, for OpenFlow (OF) based solutions \cite{calient}, 
such coupling dictates the installation of an OpenFlow rule for each detected elephant flow in order to reroute it to the OCS plane -- referred to as \textit{per-flow setup}. 
This approach results in a significant OpenFlow entry footprint \cite{mahout}\ignore{(further explained in Section \ref{subsec:configuration})}.
Furthermore, the OF rule setup rate is usually limited to tens of rules per second \cite{huang2013high}, and the OF rule installation requires \textit{outbound latency} to take effect in the data-plane. 

% Related Work -> (Flow tagging methods exists but not used in overall architecture for elephant flows offloading between EPS and OCS planes - for example: Mahout , MiceTrap)

% CiShare - overview
In this paper, we present a different approach for integrating OCS in DCN. \xxx{} inherently enables sharing of optical circuits, leading to better mice/elephant flow separation, by introducing a \emph{scalable} OpenFlow-based solution\ignore{elephant flows detection and rerouting}. 

In recent years, data-centers have been evolving towards a flatter aggregation/core hierarchy with more densely interconnected switches, also known as spine-leaf topologies.
Such topologies can deploy and adjust capacity more easily, with better manageability, and offer more deterministic network performance, particularly in latency \cite{roberts2013role}.
\xxx{} takes this trend one step further, and presents flat topology for the upper data-center packet tier by exploiting high-radix packet switches, as depicted in Figure~\ref{fig:topology}.\ignore{Figure~\ref{fig:topology} presents the concept of \xxx{} topology, in which high-radix packet switches are intra-connected by flat topology, and the OCS is connected to the packet switches.}\ignore{Any flat high-radix based topology can be used as the intra-topology between the packet switches, which is used for default routing of the flows.} The OCS is used to transmit elephant flows by creating network path shortcuts over the flat topology of the upper tier, hence dynamically allocating bandwidth between the packet switches.  
% Optical circuit sharing 
The flat upper tier topology used by \xxx{} inherently enables sharing of optical circuits by elephant flows which do not arrive from or  delivered to switches directly connected to the circuit's end-points. 

% Figure 1 example ...
Figure~\ref{fig:topology} presents an example of \textit{private optical circuit} between $S2$ and $S3$, which transmits only elephant flows that arrive from $S2$ and delivered to $S3$, through the optical circuit. 
Therefore, elephant flows from $S1$ to $S3$ are transmitted through the $S2$$\rightarrow$$S3$ packet link by $S2$\ignore{, and not through optical circuit}.
In turn, these elephant flows share the network buffers with mice flows between $S2$ and $S3$, which increases the end-to-end latency of both elephant and mice flows. 
On the other hand, the \textit{shared optical circuit} between $S6$ and $S4$ transmits any elephant flow that is delivered to $S4$ regardless its origin switch (by a corresponding $S6$ switch configuration).  
Therefore, elephant flows from $S7$ to $S4$ are transmitted through the \textit{shared} circuit by $S6$.
Hence, a better mice/elephant flow separation is obtained in the network, significantly reducing the load over the packet links between $S4$, $S5$, and $S6$, and resulting in better network performance for all flows. 
% The second, \textit{shared circuit}, is unique for \xxx{} architecture. \textit{Shared circuit} can be utilized also by flows that are transmitted through switches connected to the circuit endpoints, but originate and/or end at other switches. 
\ignore{In overall, the advantage of \xxx{} topology is two-fold: it removes the need for the expensive EPS plane used by previous work, and inherently enables optical circuits sharing.}

\xxx{} introduces SDN-based scalable elephant flow rerouting method supporting optical circuit \textit{sharing}. \xxx{} exploits the servers to detect and tag elephant flows by setting the DSCP IP field, which is usually used for packet classification. Then, the DCN orchestrator identifies the elephant flows by sampling the upper tier packet switches. Therefore, in order to redirect \textit{all} elephant flows to a given optical circuit by a packet switch, a single OF rule is required that matches the elephant flow DSCP tag and its destination. Hence, the OF rule footprint and OF flow setup rate are significantly reduced; and the outbound latency is mitigated for subsequent elephant flows after the circuit has been established. 

The contributions of this work include:\\
1) New topology concept for EPS/OCS DCN that\ignore{removes the need for EPS switch, and enable optical circuit sharing, which} further separates mice and elephant flows, thus improves network performance.\\
2) Scalable SDN-based architecture that reduces the OF rule footprint and setup rate. It also mitigates the outbound latency problem of OF switches.
\TBD{Add concise structure of the paper...}
% 
% "The end goal of this paper is TBD. To this end, ..."
% "work flow of the paper"  
\ignore{
In the paper, we answer the following questions:\\
1) In \S\ref{subsec:topology}, which topology fits for current packet and circuit switches?\\
2) In \S\ref{subsec:flow-detection}, where and how should elephant flows identified?\\
3) In \S\ref{subsec:orchestration-plane}, how to decide and configure optical circuits according to the given traffic in the network?\\
}

\section{\xxxTitle{} Topology} \label{sec:topology}
In this section, we present the concept of \xxx{} topology without delving into design and options of the upper packet tier topology. 
% OCS in-direct routing as presented in Eclipse for instance is some what different approach as compared to our work: Eclipse employs in-direct routing through the OCS plane, whereas here we leverages existing optical circuits for "combined" packet/circuit routing.

\TBD{DHPC topology offers multi-paths over the packet plane as compared to single hot spot EPS switch in hybrid solutions}

\begin{figure*}[t!]
	\centering{\includegraphics[width=0.98\linewidth]{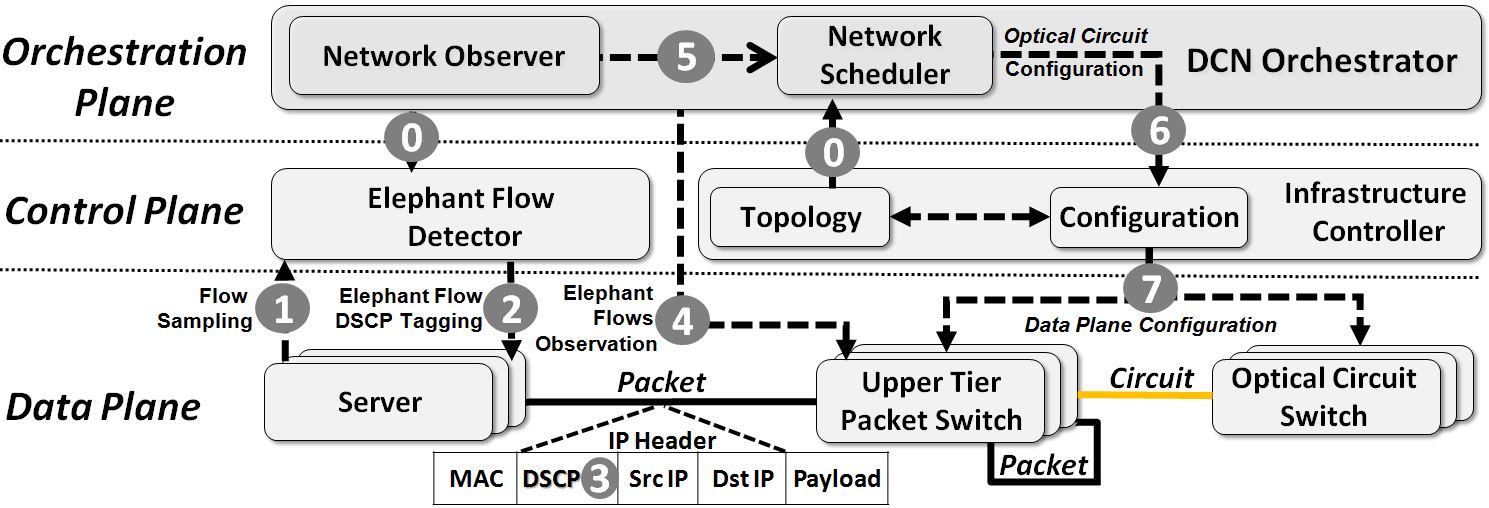}}
	\vspace{-0.1cm}
	\caption{\xxx{} architecture block diagram and workflow.}
    \vspace{-0.3cm}
	\label{fig:archoverview}
\end{figure*}

Current DCN switches offer up to 128 ports of 25\textit{Gbps} \cite{broadcom}.
In the near future, switches with 256 ports of 25\textit{Gbps} are expected and apparently will be followed by switches with 256 ports of 50\textit{Gbps}. 
As the port density increases, data-center networks become flatter with flat upper tier topology, such that the packet switches are intra-connected, thus omitting the need for an additional network layer above it. 
There are several well-known topologies, such as, multi dimensional torus or mesh, Flattened Butterfly \cite{kim2007flattened}, Dragonfly \cite{kim2008technology}, and HyperX \cite{ahn2009hyperx} that can be used to that end. 
\ignore{Our initial analysis of such topologies for the sake of intra-connectivity at the upper tier of data center topology reveals that both bi-section bandwidth, and average shortest path can be improved under proper design. }
% OCS is connected to the upper tier packet switch for shortcut and dynamic bandwidth embedding - DONE
In \xxx{} topology (Figure~\ref{fig:topology}), the OCS plane is connected to all of the packet switches at the upper  tier, and employs network path shortcuts and dynamic bandwidth allocation among them.  
% discussion about the shared optical circuits ... + figure - DONE
\ignore{The flat topology of the upper packet tier also enable better utilization of the optical circuits, thus results in better network performance.} We introduce two types of optical circuits that can be used in \xxx{} topology.

\textbf{Private Circuit} is utilized only by elephant flows that arrive from and delivered to switches directly connected to the optical circuit's endpoints, e.g. \cite{calient,farrington2011helios}.

\textbf{Shared Circuit} is inherently supported by \xxx{} topology, and can be utilized also by elephant flows that are transmitted through switches connected to the circuit's endpoints, but arrive from or delivered to other switches. 

% \ignore{The flat topology dictates that flows, which are not directly assigned to an optical circuit, should transmitted through switches at the upper packet tier Some of these flows are elephant flows that might have been transmitted through an optical circuit, but haven't due to an optical circuit conflict\ignore{ or other flows with higher QoS}. Therefore, such elephant flows are transmitted through the upper packet tier in \xxx{} (or through the EPS plane in previous approaches)}

For \textit{private} circuit configuration, elephant flows which are not assigned to an optical circuit are transmitted through the packet switches, thus might overload them. This, in turn significantly degrades the mice flows performance \cite{bowers2014advantages}. 
However, as opposed to previous solutions, \xxx{} topology dictates that some of these elephant flows are transmitted through switches which are already connected to an optical circuit.  
Therefore, by using \textit{shared} optical circuits, better mice/elephant flow separation is obtained, which results in lower congestion over the upper packet tier links, leading to better network performance. 

\section{\xxxTitle{} Architecture} \label{sec:architecture}

\ignore{
\subsection{Background}
OCS plane in data-centers is usually used for transmitting elephant flows. 
To that end, such flows have to be detected, and then rerouted to the OCS plane. 
The traditional approach of the hybrid EPS/OCS DCN monitors the electronic packet switches which connected directly to the OCS plane in order to detect elephant flows \cite{calient,farrington2011helios}. Other works \cite{wang2011cThrough} measure the traffic demand at the servers.

Such elephant flow detection requires that the monitoring method should get several samples of the same flow in order to measure its bandwidth and duration, until its detection as elephant flow. 
Then, the switch in which the flow has been detected is configured to offload the flow to the OCS plane. 
The detection and rerouting is done for each elephant flow in the network, such that a given switch must hold the configurations of all of the elephant flows that are transmitted through it (e.g., \cite{mahout}). 
Hence, such approach results in inefficiency of the flow table and therefore is not scalable.
Furthermore, the monitoring method should detect all elephant flows over a given switch, which are generated by the underlying servers; 
therefore, it imposes intensive rate monitoring requirements over the switches in order to properly detect all of the transmitted elephant flows.
}
% \subsection{\xxxTitle{} Overview}
Figure \ref{fig:archoverview} depicts block diagram of \xxx{} architecture, which\ignore{\xxx{} approach decouples the elephant flow detection, and replace it by tagging and observation processes.} decouples the elephant flow detection and rerouting phases\ignore{(e.g., \cite{calient,farrington2011helios})} to elephant flow detection and tagging, observation, and rerouting phases.
First, the egress network traffic of each server is sampled and tracked by the \textit{Elephant Flow Detector} (step 1 in Figure \ref{fig:archoverview}). 
Each flow that exceeds a given threshold for the transferred bytes and/or the flow duration (according to the criteria initialization in step 0) is detected as elephant flow, similar to \cite{al2010hedera,mahout,curtis2011devoflow}. 
Then, each detected elephant flow is tagged by setting a predefined value to the IP DSCP field\footnote{ A 6-bit field in the IP header for packet classification purposes that can be used to tag different flows types. 
For instance, \xxx{} can be extended to tag elephant flows according to different levels of bandwidth/duration thresholds, or according to different QoS\ignore{ (e.g., bandwidth or latency requirements)}.}, notated by DSCP$_{e}$ (steps 2 and 3, over the \textit{Server}\footnote{ In bare-metal based DCNs, one can tag the elephant flows by any packet modification method (e.g., by \texttt{iptables} for Linux). 
Alternately, in overlay virtualized DCNs, one can use the overlay  controller to configure the hypervisor to tag DSCP fields in the IP header encapsulation.}
and the \textit{Packet Network}, respectively).
The \textit{Upper Tier Packet Switches} (which are directly connected to the OCS plane) are monitored by the \textit{Network Observer}\ignore{ in the orchestration} plane to observe \textit{only} the tagged elephant flows and track their bandwidth and duration\footnote{ The bandwidth and duration of the tagged elephant flows can also be obtained from the \textit{Elephant Flow Detector}.} (step 4).
Studies on live DCN traffic \cite{kandula2009nature} show that elephant flows account for less than 10\% of all flows.
Therefore, tagging the elephant flows in advance by the servers and only tracking them over the packet switches significantly reduces the number of tracked flows by the \textit{Network Observer}, which reduces CPU, memory and network usage. On the contrary, detecting the elephant flows over the packet switches require significantly more network and compute resources since \emph{all} flows should be monitored. 

The \textit{Network Scheduler} decides which circuits to establish according to the current flow demand in the network (step 5)\ignore{ different criteria (e.g., aggregate offered throughput number of elephant flows between pair of packet switches)}, and informs the \textit{Infrastructure Controller} (step 6). In turn, the \textit{Infrastructure Controller} configures the data-plane accordingly (step 7). 
Then, each pair of packet switches connected to a circuit's endpoints are installed with an OF rule to reroute matched elephant flows through this circuit. The OF rule matches the DSCP$_{e}$ value in the IP header and the destination subnet connected to the switch at the other end-point of the circuit. \textit{Private circuit} is configured by matching only flows ingress from ports connected to the lower tier. \textit{Shared circuit} is configured by matching also ports connected to packet switches at the upper packet tier (section \ref{subsection:circuit_configuration}). 

% the scalability advantage 
\xxx{} architecture requires only a single flow rule in order to transmit all of the elephant flows through a given optical circuit, either \textit{shared} or \textit{private}. 
Furthermore, subsequent elephant flows, which are generated and tagged after the corresponding optical circuit has been established, are also matched by the flow rule over the packet switches to be redirected through the optical circuit. Hence, the outbound latency is mitigated, and the required OF rule footprint and OF rule setup rate are reduced (section \ref{subsec:rerouting_configuration}). 

\subsection{Private / Shared Circuit Configuration}\label{subsection:circuit_configuration}
\textit{Private} and \textit{shared} optical circuits are differed by setting which of the switch's input ports are matched by the rerouting rule of elephant flows through the optical circuit.
Therefore, different metadata values are assigned to packets from input ports connected to the lower and the upper packet tiers.
Then, by mask matching on the metadata value of an ingress packet, one can configure the switch either to use the optical circuit as \textit{private} by serving only packets from the lower tier, or \textit{shared} by serving packets also from the upper tier. 

\begin{figure}
	\includegraphics[width=0.95\linewidth]{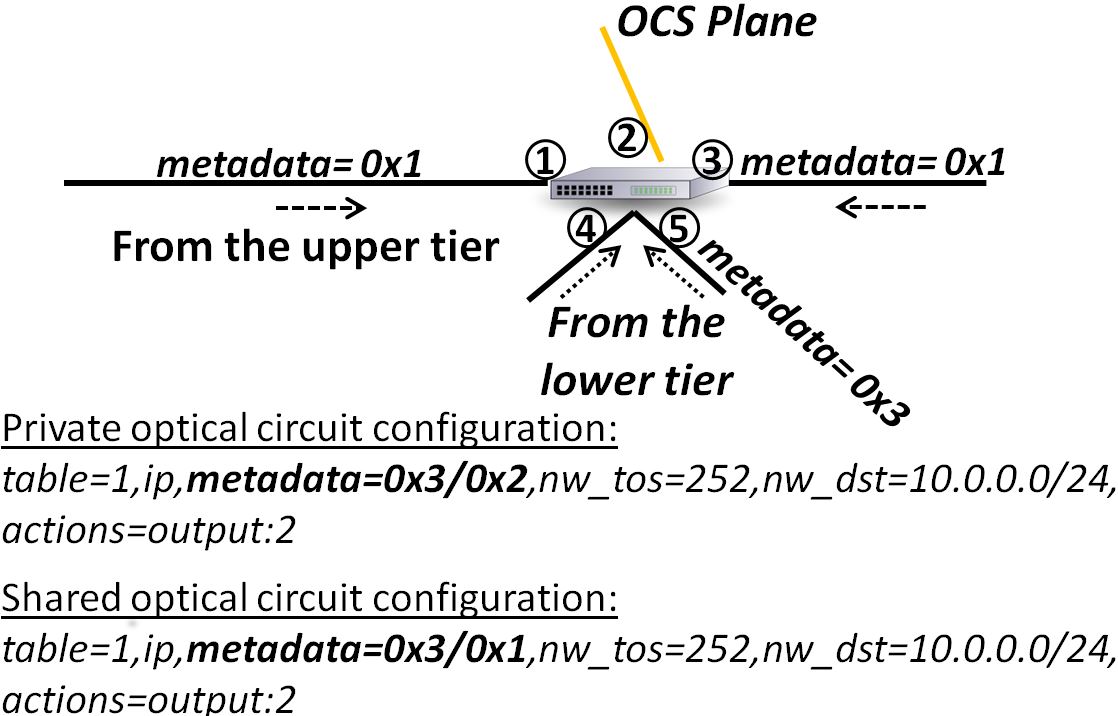}
	\vspace{-0.3cm}
	\caption{Open vSwitch example -- Private/Shared circuits configuration. A \textit{single} OF rule matches packets arriving from \textit{all} input ports of the switch connected to either lower or upper tier, by using a predefined assignment of packet metadata.}
    \vspace{-0.3cm}
	\label{fig:private_shared_circuit}
\end{figure}

Figure~\ref{fig:private_shared_circuit} demonstrates Open vSwitch \cite{pfaff2009extending} configuration for \textit{private} and \textit{shared} circuits. 
At initialization, metadata values of 0b01 and 0b11 are assigned to packets arriving from the upper and lower tier, respectively.
For a \textit{private} circuit, a \textit{single} OF rule is set to match packets with metadata values of 0b1* by using 0b10 mask. Therefore, only packets from \textit{all} input ports connected to the lower tier are matched and transmitted though the circuit. Similarity, for a \textit{shared} circuit, packets with metadata of 0b*1 are matched by using 0b01 mask. Hence, packets arriving from \textit{all} input ports connected to both upper and lower tiers are matched and transmitted through the circuit. 
As described above, the OF rule is also set to capture the DSCP$_e$ value (\textit{nw\_tos}), and the lower tier subnet destination (\textit{nw\_dst}) of the switch connected to the other end of the optical circuit.

\ignore{
\old{First at} \new{During} initialization, metadata value of $0b01$ ($0b11$) \SV{unclear. what does it mean to assign a value to a packet?} is assigned to packets from input ports connected to switches at the upper tier (lower tier).
Then, in order to configure a \textit{private} circuit, a flow descriptor is set to capture packets with metadata values 
that equals 0b1*\ignore{their second bit equals '1'}, by 0b11/0b10 masking; hence, capturing only packets from switches at the lower tier. 
Similarly, in order to configure a \textit{shared} optical circuit, packets with metadata of 0b*1 are captured, i.e., packets from both the lower tier (0b11) and the upper tier (0b01), by \SV{confusing. what does it mean 0b11/0b01 masking?} 0b11/0b01 masking.
As described above, the flow descriptor is also set to capture the DSCP$_e$ value (\textit{nw\_tos}), and the destination of the switch connected to the other side of the optical circuit. 
}
\subsection{Scalable Elephant Flow Rerouting}\label{subsec:rerouting_configuration}
% \textbf{Scalable Elephant Flow Rerouting:}\label{subsec:configuration}\\
% \textbf{Open-Flow Rules FootPrint Savings:}\\
By leveraging the DSCP tagging of the elephant flows and the packet metadata assignment according to their corresponding input ports, \xxx{} results in a single OF rule for each switch that is connected to an optical circuit's end-point. 
Therefore, \xxx{} significantly reduces the OF footprint, as compared to previous works which requires an OF rule for each rerouted flow -- \textit{per-flow setup}, e.g. \cite{calient,al2010hedera,mahout}. 

Assuming that there are on average 1k simultaneous elephant flows \cite{al2010hedera,benson2010network,kandula2009nature} between two packet switches at the upper tier, means that existing approaches require 1k OF rules for each of the packet switches, which might consume most of current OF switches flow table size. 
For instance, HP ProCurve 5400zl switches support up to 1.7K OpenFlow entries \cite{mahout}; HP ProCurve J9451A supports 1.5k OF entries \cite{huang2013high}; HP ProCurve 5406zl, Pica8 P-3290, and Dell PowerConnect 8132F support up to 1.5k, 2k and 750 rules, respectively \cite{kuzniar2015you}.
Hence, the currently used \textit{per-flow setup} approach results in average flow table consumption of 50\%-67\% for elephant flows rerouting. 
Since \xxx{} requires only single OF rule for each circuit, it result in a significantly  smaller OF footprint, as we demonstrate in our evaluation (section \ref{sec:evaluation}). 
Furthermore, OF switches have limited OF rule setup rate. For instance, \cite{huang2013high} indicates that flow rule setup rate of OF switches is limited to approximately 40 flow/sec. 
Clearly, \xxx{} significantly reduces the required OF setup rate; hence, proposes feasible solution for current OF switches.

\ignore{
In \xxx{} architecture the elephant flows are already pre-tagged at the servers.
Therefore, by leveraging the DSCP tagging of the these flows,
\xxx{} \ignore{offers an efficient rerouting configuration of the switches, which} requires only single set of OF match fields at each packet electronic switch connected to the optical circuit end-points.
For instance, in order to transmit only elephant flows through the optical circuit, the flows' match fields captures all flows that have a given DSCP$_e$ and that their destination is the switch at the other end-point of the optical circuit. 
In order to transmit also mice flows through the optical circuit (e.g., when significant number of mice flows are detected between two packet electronic switches), the flows' match fields captures flows only by their destination.
}
Once an optical circuit is configured, subsequent ingress elephant flows arriving to the packet switches are matched by the OF rule and ,in turn, transmitted through the optical circuit. 
Consequently, \xxx{} mitigates the OF outbound latency\footnote{ The latency of the switch to install/modify/delete OpenFlow rules provided by the SDN controller.} for such subsequent flows. The OF outbound latency has been measured by previous works; \cite{he2015latency} reports that the outbound latency can be as high as 30ms.  
\cite{rotsos2012oflops} measures the outbound latency of two switches by using OFLOPS. They report ranges of 50-1000ms and 8-2000ms depending on the number of inserted flow entries. 
\cite{kuzniar2015you} measures outbound latency of up to 400ms.
The outbound latency is at the same order of the 3D MEMS OCS reconfiguration penalty or even higher. 
However, the OCS reconfiguration penalty affects the network only once for each optical circuit configuration. Whereas, the outbound latency penalty has a larger network degradation potential.
Therefore, by avoiding this additional latency for each subsequent elephant flow served by an optical circuit, \xxx{} results in better network performance. 
% Furthermore, OF switches have limited OF rules setup rate. For instance, \cite{huang2013high} indicates that flow setup rate of OF switches is limited to approx. 40 flows/sec. 
% Therefore, by mitigating the requirement to install new OF rule for each new elephant flow, \xxx{} significantly improves the OF switches response and performance for new incoming flows for an existing optical circuit.  

\ignore{
\subsection{Summary}
1. Elephant flow tagging at the server can be employed by either bare metal of virtualized DCN, centrally or locally. \\
2. Pre-tagging reduces load over the network observer. \\
3. Hysteresis-based algorithms reduces unjustified optical circuit replacements.\\
4. Single Open-Flow rule is required at each switch which connected to optical circuit. Hence, mitigating the outbound Open-Flow latency, and significantly reduce the Open-Flow entry footprint. \\
}
\ignore{
The benefits: 
A) New incoming elephant flow, which its source and destination ToRs are already connected by a circuit switch, will be automatically transferred through the optical circuit. the sFlow sampling in that case are used just to update the aggregated bandwidth of the elephant flows over the given circuit. Therefore, only a single OpenFlow rule over each pf the ToRs are required in order to transmit all elephant flows between them. 
B) Furthermore, we reduce (save) the elephant flows identification (according to its DSCP value) by the ToRs.
C) Different sampling and polling rates can be configured for either detecting elephant flows over the hypervisor, and re routing them over the packet switches. This allows better and even optimized configuration for each stage.
}

\section{Evaluation}\label{sec:evaluation}

%\subsection{General Setup}
% Topologies. 
\textbf{Topologies:} we evaluate \xxx{} for two flat upper tier topologies: Ring and Flatted Butterfly \cite{kim2007flattened}.
Ring topology offers a simple wire connectivity, and is used by industrial DCNs. Facebook \cite{farrington2013facebook} presents a DCN architecture which uses Ring topology to connect the cluster and aggregation switches; and Google \cite{singh2015jupiter} uses Ring topology to connect cluster routers. 
Flattened butterfly (FBFly)  takes advantage of high-radix switches to create a scalable, yet low-diameter network. Google \cite{abts2010energy} show that FBFly is a power efficient topology for high-performance datacenter networks.
For both topologies, the bandwidth of the packet and circuit links are set for 1/10 ratio, as used by \cite{REACToR}.
% Traces
We use two DCN traces to simulate aggregated traffic to the upper packet tier, with skewed and uniform traffic patterns.

\textbf{Traces from the University of Wisconsin (\textit{UNI1})} are presented in \cite{benson2010network}, which contain recorded traffic among approximately 2900 servers for a one hour duration. 
Analysis of this trace by \cite{liuscheduling} shows mostly \textit{sparse and skewed traffic}.
We analyze \textit{UNI1} pcap traces and extract the TCP sessions properties, and their start time. Then, in order to simulate DCNs with different number of hosts, 
we consolidate the hosts by subnets, and merge the traffic for each subnet to represent a node in our modified trace. 
The subnet sizes are chosen accordingly to meet the required number of hosts. 
In addition, we reduce the time intervals between the sessions to obtain moderate network load.

\textbf{Synthetic Data Center Trace (\textit{Uniform})} is created based on traffic characteristics from \cite{al2010hedera,alizadeh2010data,kandula2009nature,REACToR}, such that elephant flows are 10\% of the number of flows and accommodate 90\% of the demand.\SV{in UNI1 traces on average mice=8-9Kb elephant=383-547Kb. In Uniform: mice=573-675Kb elephant=52-54.6Mb. This is the reason for differences in orders of magnitude in the Tables. We can safely say that the uniform trace stresses the topology. We can put the mice/elephant flow distribution in the description of the traces. There we can state that UNIFORM is much more intensive to stress the topology. I also think we can rename uniform$\rightarrow$synthetic. Since the skewed-uniform here is less important than moderate-intensive differentiation. It may be the case that the intensity of the trace plays a bigger role than skew. Also, since we target upper tier, skew should be moderate unless placement has failed.}
We generate traffic with random distribution of sessions between mice flow traffic (2KB to 32KB) and elephant flows (up to 100MB) \cite{alizadeh2010data, greenberg2009vl2}, with \textit{uniform traffic} distribution \cite{al2010hedera}.

\subsection{Emulation}\label{subsec:emulation}

\begin{figure}[t!]
\centering 
\subfigure[Average Mice Throughput] {
\includegraphics[width=0.47\textwidth]{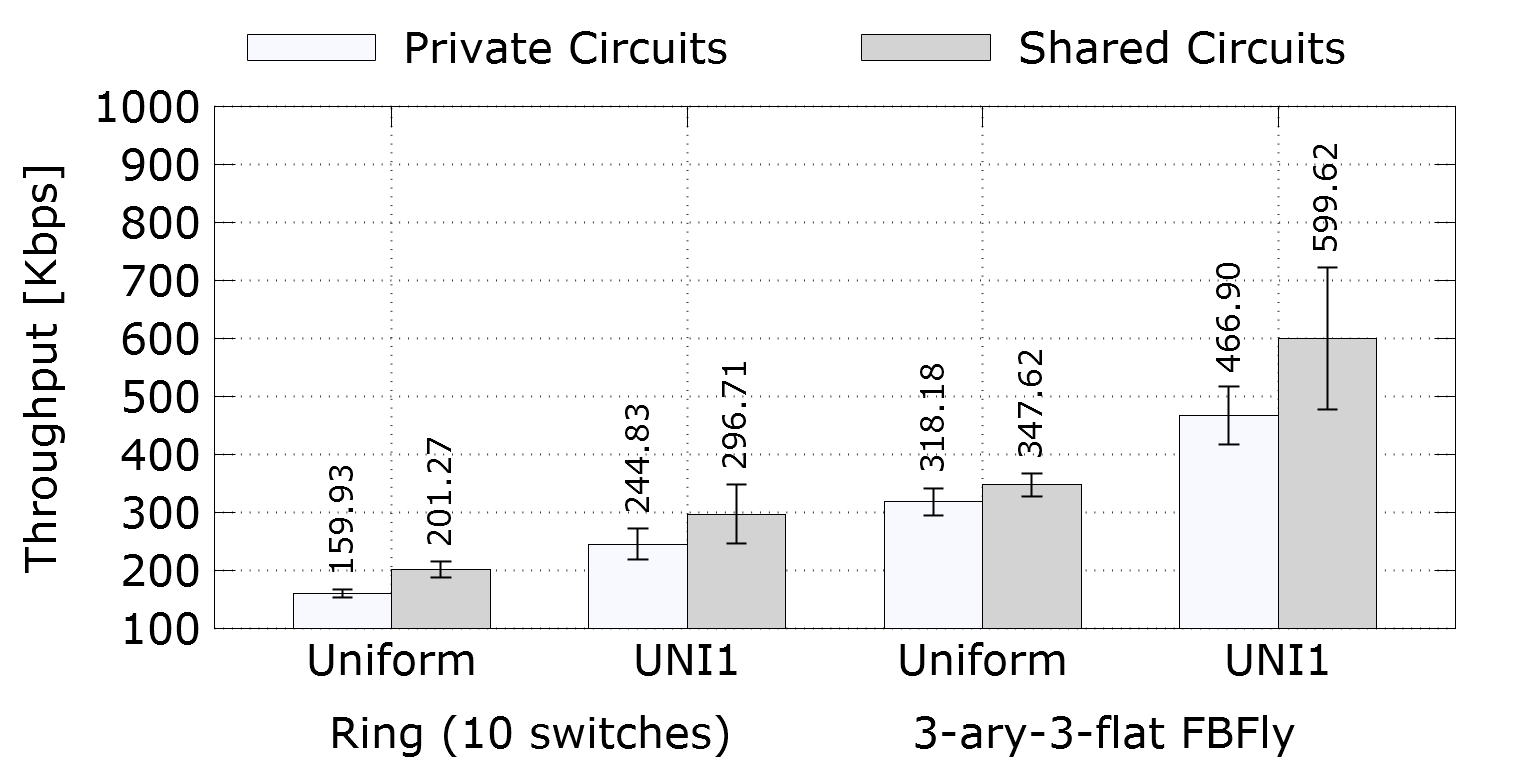}
\label{fig:Throughput_mices}}
\subfigure[Average Elephant Throughput] { 
\includegraphics[width=0.47\textwidth]{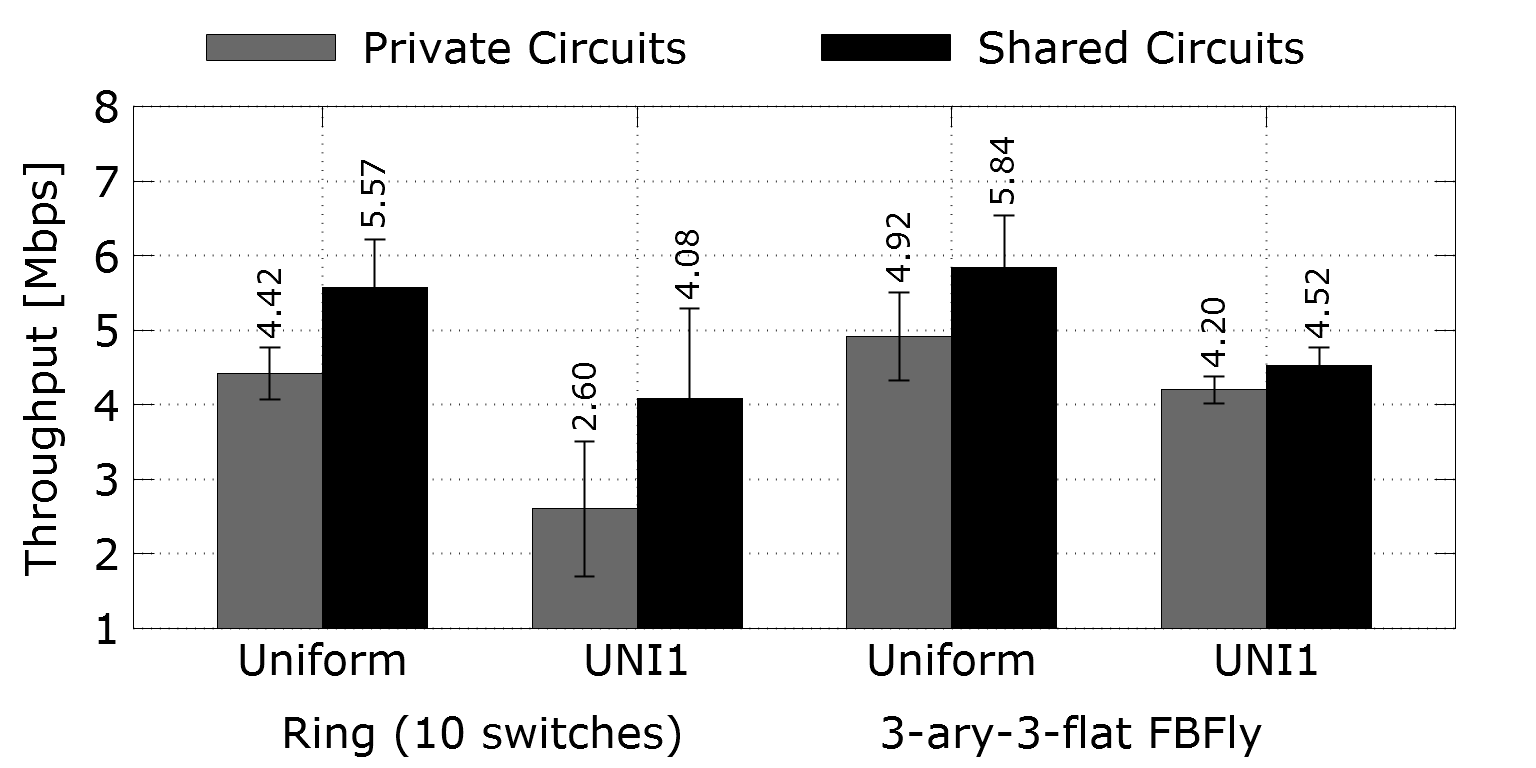}
\label{fig:Throughput_elephants}}
\vspace{-0.3cm}
\caption{Average throughput as reported by \texttt{iperf3} for Mininet environment under \textit{moderate} network load.}
\label{fig:Throughput}
\vspace{-0.3cm}
\end{figure}

We develop an emulated environment of \xxx{} by using Mininet \cite{lantz2010network} version 2.2.1 running over an IBM x3550 M4 server with 196GB of RAM, 24 Xeon-E5-2630@2.3GHz CPUs (with six cores each), and Ubuntu 14.04 with Linux 3.19 kernel.
We use sFlow \cite{wang2004sflow} to sample the egress flows of the hosts by the \textit{Elephant Flow Detector}, and to sample the Open vSwitches by the \textit{Network Observer}.
The OCS is emulated by a constrained  Open vSwitch to employ optical circuits, such that only one input port can be configured to transmit to any given output port.
Each OCS reconfiguration is emulated by first removing the colliding optical circuits, and configuring the new requested optical circuits after a 20\textit{ms} delay to emulate 3D MEMS OCS typical reconfiguration penalty, e.g. \cite{polatis}. We evaluate an upper tier Ring with 10 packet switches and 3-ary-3-flat FBFly (9 packet switches) with packet and circuit links of 10\textit{Mbps} and 100\textit{Mbps}, respectively. The network traffic is generated by \texttt{iperf3} \cite{iperf3} according to \textit{UNI1} and uniform traces configured for moderated network load without hitting the CPU-bound of the server that running Mininet. 

% traces , topologies, results, 
Figure~\ref{fig:Throughput} presents a comparison of the average throughput as reported by \texttt{iperf3} between mice and elephant flows, for both network traces over the Ring and FBFly topologies. In general, \textit{shared} circuits improve the throughput of both elephant and mice flows as compared to \textit{private} circuits.
In particular, we observe that: (1) Skewed traffic (\textit{UNI1} trace) introduces patterns which can be exploited by \textit{shared} circuits, such as many elephant flows from different sources to the same destination. Therefore, \textit{shared} circuits further improve the network performance of skewed traffic, for instance by 29\% for mice flows over FBFly, and 57\% for elephant flows over Ring; whereas, uniform traffic is improved by 9\% and 26\%, respectively. (2) The connectivity of Ring topology is limited, which results in degraded performance as compared to FBFly. Therefore, the connectivity and network throughput of Ring topology can be further improved by the \textit{shared} circuits. In particular, the \textit{shared} circuits improve the network throughput of Ring topology by 21\%-57\%, as compared to FBFly which is improved by 8\%-28\%. 
In addition, Table \ref{table:of_footprint} presents the OF rule footprint of \textit{UNI1} and \textit{uniform} traces, under moderate network load.  
\ignore{Table \ref{table:duration} presents the completion time of mice and elephant for \textit{private} and \textit{shared} circuit configurations. The \textit{shared} optical circuit configuration reduces the completion time by 15\%-20\% as compared to \textit{private} circuits configuration. 
Table \ref{table:of_footprint} presents the average OF rule footprint for one minute time slots, in order to reroute the elephant flows through the optical circuits for \textit{per-flow setup} and \xxx{} approaches. It can be seen that \xxx{} significantly reduce the OF footprint.}

\begin{table}[t!]
\small
\centering
\begin{tabular}{|c|c|c|c|c|c|}
\hline
\multicolumn{2}{|c|}{\multirow{2}{*}{\diagbox[width=12em]{\small Trace$\mid$Method}{\small Topology$\mid$Circuit}}} & \multicolumn{2}{l|}{\centering{Ring (10 Switches)}} & \multicolumn{2}{l|}{\centering{3-ary-3-flat FBFly}} \\ \cline{3-6} 
\multicolumn{2}{|l|}{}                                 & Private  & Shared & Private  & Shared       \\ \hline \hline
\multirow{2}{*}{\textit{UNI1}}    & \textit{Per-flow setup}     &     445        &   449         &   398          &  384        \\ \cline{2-6} 
                              & \xxx{}                 &     26         &  31           &   20           &  17           \\ \hline \hline
\multirow{2}{*}{\textit{Uniform}}  & \textit{Per-flow setup}      &    563  &    588    &   486     &   435     \\ \cline{2-6} 
                              & \xxx{}                 &    45   &   52      &   37      &   31     \\ \hline
\end{tabular}
\SV{This is approximation. Unfortunately can't measure this exactly. Also This is an extrapolation of the mininet results since the played trace is not long enough. For UNI1 these numbers reflect 8 seconds and for Uniform 1 min (since mininet links struggle with the big elephant flows. In practice we will probably see a much bigger advantage for \xxx{} but unfortunately we cant show it here). This is also highly sensitive to mice/elephant flow sizes. The good news are that we are better no matter what and the advantage is big. I think we just need to call it as it is: the footprint during the truncated/scaled traces played by mininet. The limitations should be clear to everyone who ever tried mininet. The simulation results show the real numbers of the full traces later.}
\vspace{-0.3cm}
\caption{OF rule footprint for elephant flow rerouting during one minute of trace, under \textit{moderate} network load. \xxx{} significantly reduces the OF footprint.}
\vspace{-0.5cm}
\label{table:of_footprint}
\end{table}

\ignore{
\begin{table}[t!]
\small
\centering 
\label{my-label}
\begin{tabular}{|c|c|c|c|c|}
\hline
\multirow{2}{*}{\diagbox[width=6.5em]{\small Method}{\small  Topology}} & \multicolumn{2}{l|}{\centering{Ring (10 Switches)}} & \multicolumn{2}{l|}{\centering{3-ary-3-flat FBFly}} \\ \cline{2-5} 
                                         &     Private      &   Shared        &    Private       &   Shared        \\ \hline
         \textit{Per-flow setup}         &     445          &   449          &   398           &  384         \\ \hline
         \xxx{}                          &     26         &  31           &   20           &  17         \\ \hline
\end{tabular}
\caption{OF rule footprint for elephant flows rerouting. \xxx{} significantly reduces the OF footprint.}
\vspace{-1.7cm}
\label{table:of_footprint}
\end{table}}

\subsection{Simulation}

\begin{figure*}[t!]
	\centering{\includegraphics[width=0.93\linewidth]{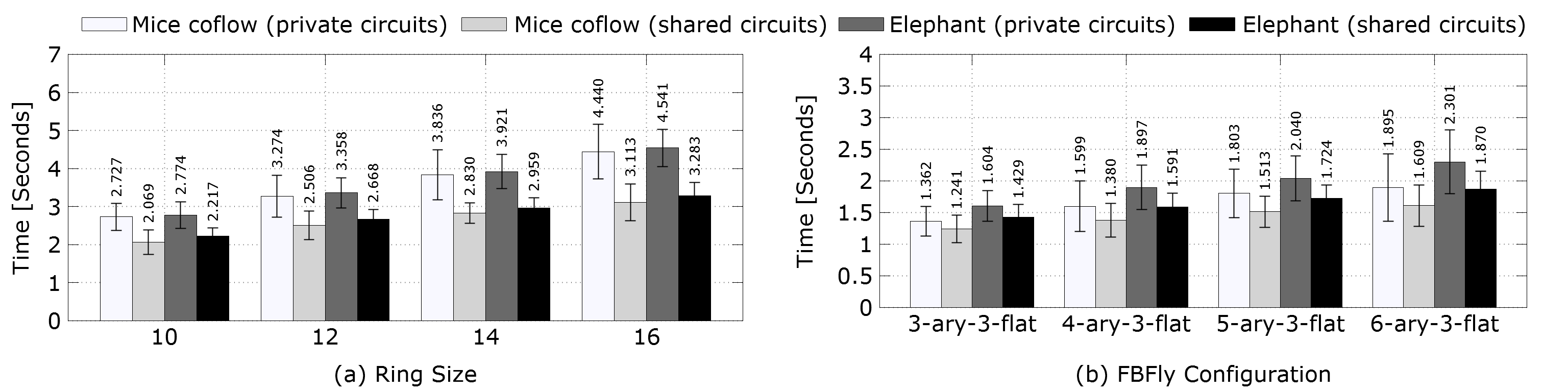}}
	\vspace{-0.4cm}
	\caption{Average completion time of mice coflows and elephant flows under \textit{intensive} network load, over two upper tier topologies: (a) Ring (10 to 16 packet switches) and (b) FBFly (9 to 36 packet switches).}
	\vspace{-0.2cm}
	\label{fig:Duration_Scalability}
\end{figure*}

% \begin{figure*}[t!] 
% \centering 
% \subfigure[Ring] {
% \includegraphics[width=0.45\textwidth]{Figures/Duration_Scalability.png}
% \label{fig:Duration_Scalability_ring}}
% %\hfill 
% \subfigure[FBFly] { 
% \includegraphics[width=0.45\textwidth]{Figures/fbfly_Duration_Scalability.png}
% \label{fig:Duration_Scalability_fbfly}}
% \vspace{-0.3cm}
% \caption{Average completion time for mice and elephant flows in (a) Ring and (b) Flattened Butterfly Topologies.}
% \vspace{-0.3cm}
% \label{fig:Duration_scalability}
% \end{figure*}

\begin{figure*}[t!]
	\centering{\includegraphics[width=0.93\linewidth]{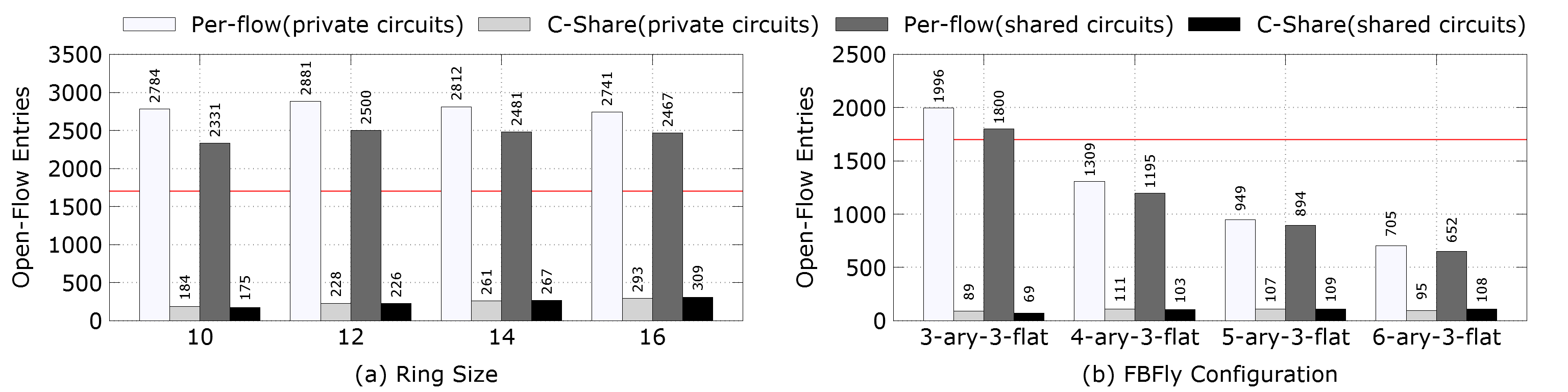}}
	\vspace{-0.4cm}
	\caption{OpenFlow rule footprint per switch during one minute of network trace under \textit{intensive} network load, \qquad over two upper tier topologies: (a) Ring (10 to 16 packet switches) and (b) FBFly (9 to 36 packet switches). \qquad \qquad \qquad The horizontal line indicates 1.7k OF rule entries. Any OF entries count above it might be an unfeasible scenario.}
	\vspace{-0.4cm}
	\label{fig:of_count}
\end{figure*}

% \begin{figure*}[t!] 
% \centering 
% \subfigure[Ring] {
% \includegraphics[width=0.45\textwidth]{Figures/ring_of_rules_count.png}
% \label{fig:of_count_Scalability_ring}}
% %\hfill 
% \subfigure[FBFly] { 
% \includegraphics[width=0.45\textwidth]{Figures/fbfly_of_rules_count.png}
% \label{fig:of_count_Scalability_fbfly}}
% \vspace{-0.3cm}
% \caption{OF rules footprint per packet switch during one minute. For two upper tier topologies: (a) Ring and (b) FBFly topologies. The red-line indicates 1.7k OF rule entries.}
% \TBD{x-axis title: FBFly Configurations. legend: try Per-flow setup (private circuit) . change to y-title to "OpenFlow Entries" . Ring Size / FBFly Configuration. -- Unfeasible for \textit{per-flow setup}}
% \vspace{-0.3cm}
% \label{fig:of_count}
% \end{figure*}

We use an event driven simulation to evaluate the completion time of mice coflows and elephant flows, and measure the corresponding OF rule footprint for rerouting elephant flows by the packet switches through \textit{private} or \textit{shared} circuits. 
We use the synthetic \textit{uniform} traces to demonstrate the scalability of \xxx{}  under \textit{intensive} network load. Specifically, we generate network traffic comprised of mice coflows and elephant flows. The mice coflows are 90\% of the number of flows, and accommodate 10\% of the total demand. 
We simulate Ring and Flattened Butterfly upper packet tier topologies, with varied number of packet switches. Each packet switch serves 40 hosts.
The packet and circuit links are set to  10\textit{Gbps} and 100\textit{Gbps}, respectively.

Figure \ref{fig:Duration_Scalability} presents the average completion run-time of mice coflows and elephant flows for \textit{private} and \textit{shared} circuit configurations over 60 trials. 
The \textit{shared} circuits improve the average completion time by 20\% for a Ring with 10 switches and up to 30\% for a Ring with 16 switches.
The Ring topology is unscalable in terms of connectivity.
Therefore, the \textit{shared} circuits can significantly increase the topology connectivity and mice/elephant flow separation, which results in increased improvement of the completion time as the Ring size increases.
On the other hand, since FBFly is scalable, the improvement of the completion time by \textit{shared} circuits equals 15\%-20\% for all FBFly sizes; hence, the \textit{shared} circuits results in relatively constant mice/elephant flow separation degree.
The same applies for the OF rule footprint presented in Figure \ref{fig:of_count}. The required OF rules of \textit{per-flow setup} for Ring remains constant and higher than 1.7k (prevalent OF table size \cite{mahout,huang2013high,kuzniar2015you}). On the other hand, due to the scalability of FBFly, as the size of FBFly increases, less OF rules are required for rerouting the elephant flows through  \textit{private} or \textit{shared} circuits. However, the OF footprint of \textit{per-flow setup} is still high and is significantly reduced by \xxx{}.

\section{Related Work} \label{sec:related-work}
% in-direct routing ... - however never applied to current state of the art solutions

% mice aggregation - orthogonal solution for shared optical circuits, however more complex, and less reactive to the fast varying nature of mice flows ...

%Helios, cThrough, REACTOR,  what else? I believe these 3 are the closest to DHPC.

%Some of the indivdual ideas used in the current work have been discussed in the literature however maybe in a different architecture. For instance scheduling of packets, introduction of optics in DCs, etc. They deserve to be credited in our work, however, this can be done while explaining the architecture and algorithms and not necessarily devote a section to them.
\ignore{
\textbf{Hybrid Packet/Circuit DCN Architectures:}\\
}
\ignore{
Several hybrid DCN architectures for electronic packet and optical circuit switches have been proposed. 
To the best of our knowledge, all of these previous works (e.g.,\cite{farrington2011helios,wang2011cThrough,calient}) are based on separated EPS and OCS planes, in which EPS is used for fast all-to-all communication between the switches, while OCS is used for high-bandwidth, slowly varying, and usually long-lived communication between the switches.}

EPS/OCS DCN solutions, e.g. \cite{farrington2011helios,REACToR,wang2011cThrough}, present different approaches for integrating OCS in DCN. The control planes presented in these works are based on non-SDN methods, thus limited as compared to SDN-based solutions. 
c-Through \cite{wang2011cThrough} uses predefined VLANs for static EPS/OCS planes, and tags elephant flows with the corresponding VLAN, without the ability to dynamically configure the network. 
Helios \cite{farrington2011helios} implementation consists of Monaco packet switches and sets its forwarding table to reroute all flows that are delivered to a specific destination pod, without the ability to separate among mice and elephant flows to EPS and OCS planes, respectively.
REACToR \cite{REACToR} presents state-of-the-art FPGA-based solution; however, do not propose an elephant tagging and rerouting technique. Furthermore, these works are based on separated EPS and OCS planes, which restrict the seperation of mice and elephant flows in the network. 
On the other hand, \xxx{} inherently supports such mice/elephant flow separation. 

ProjecToR \cite{ghobadi2016projector} presents a free-space optics (FSO) solution for DCN, composed of \textit{dedicated} and \textit{opportunistic} optical links. \xxx{} can be employed over such solution, and offer optical circuit sharing over the \textit{dedicated} optical links. 

SDN-based works present elephant \cite{mahout} and network-limited \cite{al2010hedera} flow scheduling for EPS-only DCNs.~\cite{calient} presents SDN-based solution for OCS/EPS DCN. These works use a specific OF rule for each rerouted flow. Hence, they introduce the aforementioned OF scalability issues. Namely, table rule footprint, setup rate, and outbound latency. All are mitigated by \xxx{}.

\section{Conclusions and Future Work} \label{sec:conclusions}

In this paper we proposed \xxx{}, a new approach for integrating OCS in DCN. We demonstrated how \xxx{} inherently supports circuit sharing that further separates mice and elephant flows leading to increased performance. We presented a scalable SDN-based solution including elephant flow rerouting that requires only a single OF rule per circuit. 

This work is a starting point on a way towards a full-fledged implementation of \xxx{}. We list two of \xxx{} advanced architectural aspects.

\textbf{Advanced Circuit Sharing:} We presented \textit{shared} circuits only for \emph{last hop routing}. Namely, the sharing is employed for elephant flows delivered to one of the circuit's endpoints. By advanced configuration, we can enable circuit sharing with elephant flows at any hop along their routes.

\textbf{Upper Tier Topology:} \xxx{} is evaluated for Ring and FBFly upper tiers. Other topologies might offer better mice/elephant flow separation by exploiting circuit sharing.

\ignore{
First, elephant flows that are delivered only to the connected optical circuit's end-points can utilize the optical circuit sharing; hence, the \textit{shared} circuits are used only for \emph{last-hop} routing. However, by advanced configuration, the circuits can be shared at any hop of the elephant flows route path, further increasing the mice/elephant flow separation. 
Second, \xxx{} is evaluated for Ring and FBFly; however, other topologies might offer better circuit sharing support.} 

\ignore{
Better shared optical circuits usage, which employ elephant flows 
Shared optical circuits usage - not just as last-hop towards destination, but also at the beginning/middle of the flow's route path 
Private/Shared optical circuits - dynamic decision
The orchestrator can decide whether to configure private/shared circuit in order to maintain its utilization according to current elephant flows bandwidth, and desired QoS.
Topology design - intra-connectivity, number of optical circuits, etc ..}

% \section*{Acknowledgments}

\bibliographystyle{abbrv} 
\begin{small}
\bibliography{ref}
\end{small}
\label{last-page}

\end{document}